\documentclass[10pt]{article}

\usepackage{amsmath}
\usepackage{amssymb}  
\usepackage[T1]{fontenc}
\usepackage{pifont}
\usepackage{pstricks}
\usepackage{amsfonts}
\usepackage{graphicx}
\usepackage{amsmath}
\usepackage{authblk}
\usepackage{pstricks} 
\usepackage{pstricks-add}
\usepackage{caption}
\usepackage{placeins}
\usepackage{pstricks}
\usepackage{pdfpages}
\usepackage{framed}
\usepackage{cancel}
\usepackage{bm}
\usepackage{epstopdf}
\def\rr2dot{\mathop{\bf r}\limits}
\def\x2dot{\mathop{x}\limits}
\def\y2dot{\mathop{y}\limits}

\def\bfy2dot{\mathop{\bf y}\limits}
\def\z2dot{\mathop{z}\limits}

\def\csi2dot{\mathop{\xi}\limits}
\def\et2dot{\mathop{\eta}\limits}
\def\bet2dot{\mathop{\beta}\limits}
\def\t2dot{\mathop{\theta}\limits}
\def\s2dot{\mathop{\sigma}\limits}
\def\d2dot{\mathop{\delta}\limits}
\def\q2dot{\mathop{q}\limits}
\def\l2dot{\mathop{\lambda}\limits}
\def\ps2dot{\mathop{{\cal E}}\limits}
\def\tet2dot{\mathop{\theta}\limits}
\def\bfx2dot{\mathop{\bf x}\limits}
\def\bfy2dot{\mathop{\bf y}\limits}
\def\bfq2dot{\mathop{\bf q}\limits}
\def\bbfq2dot{\mathop{\bar {\bf q}}\limits}
\def\w2{\mathop{W}\limits}
\def\xgrande2dot{\mathop{\bf X}\limits}
\def\p02dot{\mathop{P}\limits}
\def\a2dot{\mathop{A}\limits}

\newtheorem{prop}{Proposition}
\newtheorem{propr}{Property}
\newtheorem{rem}{Remark}
\newtheorem{exe}{Example}

\title{${\check {\rm C}}$etaev condition for nonlinear nonholonomic systems and homogeneous constraints}
\author{F.~Talamucci}
\affil{{\it DIMAI, Dipartimento di Matematica e Informatica ``Ulisse Dini''},\\
{\it	Universit\`a degli Studi di Firenze, Italy}\\
{\it	e-mail: federico.talamucci@unifi.it}}
\date{}

\begin{document}
	\bibliographystyle{plain}
	
	\setcounter{equation}{0}

	\maketitle
	
	\vspace{.5truecm}
	
	
	\vspace{.5truecm}
	
	
	\vspace{.5truecm}
	

\begin{abstract}
	
	\noindent
	We first present a way to formulate the equations of motion for a nonholonomic system with nonlinear constraints with respect to the velocities. The formulation is based on the ${\check {\rm C}}$etaev condition which aims to extend the practical method of virtual displacements from the holonomic case to the nonlinear nonholonomic one.
	The condition may appear in a certain sense artificial and motivated only to coherently generalize that concerning the holonomic case.
	In the second part we show that for a specific category of nonholonomic constraints (homogeneous functions with respect to the generalized velocities) the ${\check {\rm C}}$etaev condition reveals the same physical meaning that emerges in systems with holonomic constraints. In particular the aspect of the mechanical energy associable to the system is analysed.

\end{abstract}

\section{Introduction}

\noindent
The vast and growing interest in systems with kinematic constraints is motivated by a series of significant applications, among which we quote the motion of vehicles.
From a modelling point of view it is a question of extending the consolidated procedure inherent to holonomic systems (i.~e.~those subject to geometric constraints which concern the spatial coordinates only) by admitting the presence of restrictions on the velocities (kinematic constraints) and in some cases even more complex limitations.
The transition from the linear case (in some way similar to the holonomic case), namely the linear dependence of the constraints equations on the velocities, to that of nonlinear kinematic constraints
considerably complicates the treatment from a formal point of view.
If on the one hand the treatment of linear nonholonomic systems dates back to over a century ago 
(for an exhaustive historical review, see the fundamental text \cite{neimark} and the more recent review \cite{leon1}), on the other the formal study of the nonlinear ones is recent and not definitive: even the  question of the actual implementation of such constraints, starting from the Appell--Hamel example \cite{appell}, is debated (\cite{neimark}, \cite{zekovic1}).

\noindent
In a very generic and not drastic way we can divide the mathematical approaches to nonholonomic systems into three categories: a first method that is certainly expanding and relies on high--profile publications (\cite{bloch}, \cite{leon2}, \cite{massa}, \cite{marle}, \cite{ranada}, \cite{saleh}) is based on differential geometry and extends the basic theory of holonomic systems to more complex concepts (jet manifolds, jet bundles).

\noindent
A second method takes advantage of the possibility of deriving the equations of motion from a variational principle (\cite{rumy}, \cite{papa}), likewise the Hamilton principle in the holonomic case; as far as we know the situation is still open and more than one publication warns about some critical aspects regarding this procedure (\cite{cron}, \cite{flan}). 

\noindent
The third approach we mention, which is the one we will develop in the paper, takes into account the virtual displacements (\cite{li} and again \cite{papa} for a comprehensive analysis of the various types of displacements) associable to the constrained system, as well as the directions tangential to the configuration space establish a way to write a set of independent equations in the range of holonomic systems.
A nonholonomic system with linear kinematic constraints can be treated essentially as a system with geometric constraints, by considering a linear subspace of the admitted directions.

\noindent
However, for nonlinear kinematic constraints a plain geometric description of the virtual displacements does not exist: a frequently accepted hypothesis for constituting the set of virtual displacements is the so-called ${\check {\rm C}}$etaev condition (\cite{cetaev}, \cite{rumycet}), which it is an accepted axiom
leading to the expected equations of motion but does not present a real justification from the laws of mechanics, as far as we can understand (\cite{li}).

\noindent
The first part of the work (Section 1) is just right dedicated to the modeling formulation the ${\check {\rm C}}$etaev condition, from which the equations for nonholonomic nonlinear constrained systems are derived. 
A properly combination of the equations lead to an information about the mechanical energy of the system (as well as in the holonomic case occurs) and the energy equation is presented in the same Section.

\noindent
In the second part (Section 2) the intention is to highlight how the validity of a property of the constraint equations (namely (\ref{baralphaalpha})) entails the removal of discrepancies that emerge between the mathematical feature of the ${\check {\rm C}}$etaev condition and the physical reading associated with it: actually, the virtual displacements prescribed by the condition play in this case the same role as the velocities consistent with the instantaneous configurations of the system, as well as it occurs for holonomic
systems.
From the mathematical point of view, it is shown that the above ``conciliant'' condition is satisfied whenever the constraints are homogeneous equations with respect to the generalized velocities; this is the central point of the work.
Even if the few examples described here do not cover the multiplicity of nonholonomic models, it is true that 
a large part of them are implemented by conditions showing homogeneity in the kinetic variable: this motivates the final consideration of Section 3, where the question is reversed, that is whether the natural condition (\ref{baralphaalpha}) is an exclusive prerogative of those systems where the nonholonomic constraints can be formulated via homogeneous equations.

\subsection{The equations of motion}

\noindent
Let us consider a mechanical system whose configurations are determined by $n$ the lagrangian coordinates $q_1$, $\dots$, $q_n$; we recall the Lagrangian equations of motion 

\begin{equation}
	\label{eqlagrt}
	\dfrac{d}{dt}\dfrac{\partial T}{\partial {\dot q}_j}-\dfrac{\partial T}{\partial q_j}={\cal F}^{(j)}+
	{\cal R}^{(j)} \qquad j=1, \dots, n
\end{equation}
where $T(q_1, \dots, q_n, {\dot q}_1, \dots, {\dot q}_n,t)$ is the kinetic energy, ${\cal F}^{(j)}$, 
$j=1, \dots, n$ the generalized forces acting on the system and  ${\cal R}^{(j)}$ the constraints forces due to the action of the $k<n$ kinematic constraints

\begin{equation}
	\label{constr}
		\phi_\nu(q_1,\dots, q_n, {\dot q}_1, \dots, {\dot q}_n, t)=0, \qquad \nu=1, \dots, k
\end{equation}
The constraints are independent with respect to the kinematic variables,
in the sense that the rank of the $k\times n$ matrix is full:
\begin{equation}
	\label{rank}
rank\,\left(\dfrac{\partial \phi_\nu}{\partial {\dot q}_j}\right)_{\footnotesize{
\begin{array}{l} \nu=1, \dots, k\\j=1, \dots, n \end{array}}}=k
\end{equation} 

\noindent
We can assume, since it is marginal in this discussion, that the active forces admit a potential $U$:
\begin{equation}
	\label{potu}
	{\mathcal F}^{(j)}=\dfrac{\partial }{\partial q_j}U(q_1, \dots, q_n,t), \quad j=1, \dots, n\quad 
\end{equation}
Hence the equations (\ref{eqlagrt}) take the form 
 \begin{equation}
 	\label{eqlagrl}
 	\dfrac{d}{dt}\dfrac{\partial {\cal L}}{\partial {\dot q}_j}-\dfrac{\partial {\cal L}}{\partial q_j}=
 	{\cal R}^{(j)} \qquad j=1, \dots, n.
 \end{equation}
The system of $n+k$ equations (\ref{eqlagrl}), (\ref{constr}) contain the $2n$ unknown functions $q_j$ and ${\cal R}^{(j)}$, $j=1, \dots, n$.

\noindent
At this point, we outline two different procedures to deal with the system:
\begin{itemize}
	\item[$(i)$] to formulate an hypothesis for the constraints forces based on the method of lagrangian multipliers,
\item[$(ii)$] to identify the set of virtual displacements similarly to the case of holonomic constraints, that is to provide the admitted variations $\delta q_1$, $\dots$, $\delta q_n$ of the coordinates  consistent with the restrictions (\ref{constr}) (virtual variations).
\end{itemize}
In the first case one sets
	\begin{equation}
	\label{multipl}
	{\cal R}^{(j)}=
	\sum\limits_{\nu=1}^k \lambda_\nu \dfrac{\partial \phi_\nu}{\partial {\dot q}_j}, \qquad j=1, \dots, n
\end{equation}
where $\lambda_\nu$, $\nu=1, \dots, k$, are unknown coefficients; in this way the unknown quantities equalize the number of equations.
Concerning $(ii)$, a generally accepted prospect is to define the virtual displacements as those which verify 
\begin{equation}
	\label{cetaev}
	\sum\limits_{i=1}^n\dfrac{\partial \phi_\nu}{\partial {\dot q}_i}\delta q_i=0, \qquad i=1, \dots, \nu
\end{equation}
known as ${\check {\rm C}}$etaev condition; the virtual work of the constraints forces along the displacements verifying (\ref{cetaev}) is assumed to be null:
$$
\sum\limits_{i=1}^n {\cal R}^{(i)}\delta q_i =0.
$$
The two approaches have in common the fact that the virtual 
work of the forces (\ref{multipl}) is zero whenever (\ref{cetaev}) holds:
\begin{equation}
	\label{riphinu}
\sum\limits_{i=1}^n {\cal R}^{(i)}\delta q_i =
\sum\limits_{\nu=1}^k \sum\limits_{i=1}^n \lambda_\nu \dfrac{\partial \phi_\nu}{\partial {\dot q}_i}\delta q_i=0
\end{equation}
Nevertheless, the starting point (\ref{cetaev}) yields an useful method which has the advantage of not making the multipliers $\lambda_\nu$ appear and of selecting a set of independent kinetic variables: actually, 
we see that condition (\ref{rank}) makes possible the explicit writing of 
conditions (\ref{constr}) in the following way (without losing generality, except for re-enumerating the variables):
\begin{equation}
\label{constrexpl}
{\dot q}_{m+\nu}=\alpha_1(q_1, \dots, q_n, {\dot q}_1, \dots, {\dot q}_m, t)\qquad \nu=1, \dots, k
\end{equation}
with $m=n-k$. The parameters ${\dot q}_r$, $r=1, \dots, m$, play the role of independent kinetic variables, and correspondingly the virtual displacements $\delta q_1$, $\dots$, $\delta q_m$ can be considered as independent; in turn, (\ref{constrexpl}) allows to express the dependent displacements as
\begin{equation}
	\label{virtdispdip}
	\delta q_{m+\nu}=\sum\limits_{r=1}^m \dfrac{\partial \alpha_\nu}{\partial {\dot q}_r}\delta q_r, \qquad \nu=1,\dots, k
\end{equation}
To verify it, we use the role of ${\dot q}_r$, $r=1, \dots, m$ as independent variables, differentiate (\ref{constr}) and write
$$
\dfrac{\partial \phi_\nu}{\partial {\dot q}_r}+\sum\limits_{\mu=1}^k \dfrac{\partial \phi_\nu}{\partial {\dot q}_{m+\nu}}\dfrac{\partial \alpha_\nu}{\partial {\dot q}_r}=0\quad \Longrightarrow \quad 
\sum\limits_{r=1}^m 
\dfrac{\partial \phi_\nu}{\partial {\dot q}_r}\delta q_r 
+\sum\limits_{\mu=1}^k\dfrac{\partial \phi_\nu}{\partial {\dot q}_{m+\mu}}
\underbrace{\sum\limits_{r=1}^m\dfrac{\partial \alpha_\mu}{\partial {\dot q}_r}\delta q_r}_{\delta q_{m+\mu}}
=0, \quad \nu=1, \dots, k.$$

\begin{rem}
Having in mind a comparison with the holonomic case and referring to the case of $N$ vectors ${\bf r}_i(q_1, \dots, q_n, t)$, $i=1, \dots, N$ locating the points of the system, assumption (\ref{cetaev}) means that the customary condition 
$$
\delta {\bf r}_i = \sum\limits_{j=1}^n \dfrac{\partial {\bf r}_i}{\partial  q_j}\delta q_j, \quad i=1, \dots, N
$$
holding for a holonomic system is replaced in the case of kinematic constraints (\ref{constr}) by
\begin{equation}
	\label{deltar}
	\delta {\bf r}_i = \sum\limits_{j=1}^n \dfrac{\partial {\dot {\bf r}}_i}{\partial  {\dot q}_j}\delta q_j, \quad i=1, \dots, N.
\end{equation}
\end{rem}

\begin{rem}
For linear kinematic constraints 
$$
\sum\limits_{i=1}^n\sigma_{\nu,i}(q_1,\dots, q_n,t){\dot q}_i+\zeta_1(q_1, \dots, \dots, q_n, t)=0, \quad \nu=1, \dots, k
$$
admitting the explicit expressions
\begin{equation}
\label{constrexpllin}
{\dot q}_{m+\nu}=\sum\limits_{j=1}^m \alpha_{\nu,j}(q_1, \dots, q_n, t){\dot q}_j+\beta_\nu(q_1, \dots, q_n,t) , \quad \nu=1,\dots,k
\end{equation}
the condition (\ref{virtdispdip}) reduces to
$$
\delta q_{m+\nu}=\sum\limits_{r=1}^m \alpha_{\nu,r}(q_1, \dots, q_n, t)\delta q_r, \qquad \nu=1,\dots, k
$$
and corresponds to the usual assumption on virtual variations adopted in texts dealing with nonholonomic theory for linear kinematic constraints (\cite{neimark}).
\end{rem}

\noindent
The approach (\ref{virtdispdip}) requires to express coherently the Lagrangian function ${\cal L}$ by means of the only independent kinetic variables:
\begin{equation}
	\label{lstar}
	{\cal L}^*(q_1,\dots, q_n, {\dot q}_1, \dots, {\dot q}_m, t)
	={\cal L}(q_1, \dots, q_n, {\dot q}_1, \dots, {\dot q}_m, \alpha_1(\cdot), \dots, 
	\alpha_k(\cdot), t)
\end{equation}
where $\alpha_\nu (\cdot)$ stands for $\alpha_\nu(q_1, \dots, q_n, {\dot q}_1,\dots, {\dot q}_n,t)$, $\nu=1, \dots, k$. In terms of ${\cal L}^*$ the equations of motion calculated starting from (\ref{eqlagrt}) and considered along the virtual displacements (\ref{cetaev}), (\ref{virtdispdip}) take the form (see \cite{tal} for details)

\begin{equation}
\label{vnl}
\dfrac{d}{dt}\dfrac{\partial {\cal L}^*}{\partial {\dot q}_r}-\dfrac{\partial {\cal L}^*}{\partial q_r}
-\sum\limits_{\nu=1}^k\dfrac{\partial {\cal L}^*}{\partial q_{m+\nu}}\dfrac{\partial \alpha_\nu}{\partial {\dot q_r}}
-\sum\limits_{\nu=1}^k  B_r^\nu \dfrac{\partial T}{\partial {\dot q}_{m+\nu}}=0
\end{equation}
for $r=1,\dots$, where
\begin{equation}
	\label{b}
	B_r^\nu(q_1,\dots, q_n, {\dot q}_1, \dots, {\dot q}_m,t)= 
	\dfrac{d}{dt}\left( \dfrac{\partial \alpha_\nu}{\partial {\dot q}_r}\right)- 
	\dfrac{\partial \alpha_\nu}{\partial q_r}-\sum\limits_{\mu=1}^k 
	\dfrac{\partial \alpha_\mu}{\partial {\dot q}_r}
	\dfrac{\partial \alpha_\nu}{\partial q_{m+\mu}}
\end{equation}
for $r=1, \dots, m$ and $\nu=1, \dots, k$. They represent an alternative formal way (without multipliers) to the set (\ref{eqlagrl}), (\ref{multipl}), (\ref{constr}). The linear case (\ref{constrexpllin}) refers to the Voronec equations for nonholonomic systems with linear kinematic constraints (\cite{neimark}).

\noindent
The unknown functions in (\ref{vnl}) are $q_1$, $\dots$, $q_n$ but only the derivatives ${\dot q}_r$, ${\q2dot^{..}}_r$ $r=1, \dots, m$ are present, owing to (\ref{constrexpl}), once the va\-ria\-bles $({\dot q}_{k+1}, \dots, {\dot q}_n)$ that are present in $\frac{\partial T}{\partial {\dot q}_{m+\nu}}$ of (\ref{vnl}) have been expressed in terms of $(q_1,\dots, q_n,$ ${\dot q}_1, \dots, {\dot q}_m,t)$. 
Evidently, the case of merely holonomic constraints which corresponds to suppress all the terms containing the functions $\alpha_\nu$ brings back to the ordinary Euler--Lagrange equations of motion for ${\cal L}^*={\cal L}$.

\begin{rem}
The coefficients in (\ref{vnl}) can be expressed in terms of the original functions $\phi_\nu$ of (\ref{constr}) by making use of the relations 
$$
\dfrac{\partial \phi_\nu}{\partial q_i}+\sum\limits_{\mu=1}^k \dfrac{\partial \phi_\nu}{\partial {\dot q}_{m+\mu}} \dfrac{\partial \alpha_\mu}{\partial q_i}=0, \quad 
\dfrac{\partial \phi_\nu}{\partial {\dot q}_r}+\sum\limits_{\mu=1}^k \dfrac{\partial \phi_\nu}{\partial {\dot q}_{m+\mu}}\dfrac{\partial \alpha_\mu}{\partial {\dot q}_r}=0
$$
holding for any $\nu=1, \dots, k$, $i=1, \dots, n$, $r=1, \dots, m$. This means that the structure of the equations does not depend on the choice (\ref{constrexpl}) for making explicit (\ref{constr}).
\end{rem}

\subsection{An equation for the energy of the system}

\noindent
We recall the Hamiltonian function
\begin{equation}
	\label{e}
	{\cal E}(q_1, \dots, q_n, {\dot q}_1, \dots, {\dot q}_n,t)=
	\sum\limits_{i=1}^n {\dot q}_i \dfrac{\partial {\cal L}}{\partial {\dot q}_i}-{\cal L}
\end{equation}
and we ascribe to it the role of energy of the system.
At the same time, we are interested in the properties of the energy function formulated by only the independent kinetic variables, that is
\begin{equation}
	\label{estar}
	{\cal E}^*(q_1, \dots, q_n, {\dot q}_1, \dots, {\dot q}_m,t)= \sum\limits_{r=1}^m {\dot q}_r 
	\dfrac{\partial {\cal L}^*}{\partial {\dot q}_r}-{\cal L}^*.
\end{equation} 
We can say that (\ref{e}) is the energy that is naturally associated to (\ref{eqlagrl}) with the Lagrangian ${\cal L}$, while (\ref{estar}) is spontaneously connected to (\ref{vnl}), where only the independent velocities are present.
Although in a much more complex formal context of geometric type, the two notions of energy can be traced back to those present in \cite{fasso}.

\noindent
Even if we express (\ref{e}) and (\ref{estar}) by means of the same set of variables, replacing in the first one the dependent velocities ${\dot q}_{m+\nu}$, $\nu=1, \dots, k$ by virtue of (\ref{constrexpl}), we do not find the same function: actually, it can be seen without difficulty that the following relation occurs:
\begin{equation}
	\label{rele}
	{\cal E}^*={\cal E}+
	\sum\limits_{\nu=1}^k({\overline \alpha}_\nu-\alpha_\nu)\dfrac{\partial {\cal L}}{\partial {\dot q}_{m+\nu}}
\end{equation}
where we set 
\begin{equation}
	\label{baralpha}
	{\overline \alpha}_\nu (q_1, \dots, q_n, {\dot q}_1, \dots, {\dot q}_m,t)=
	\sum\limits_{r=1}^m \dfrac{\partial \alpha_\nu}{\partial {\dot q}_r}{\dot q}_r, \quad \nu=1, \dots, k
\end{equation}
At the same time, it is possible to achieve the two equations that ${\cal E}$ and ${\cal E}^*$ verify, coming from (\ref{eqlagrl}) and (\ref{vnl}) by means of the standard procedure of multiplying by the generalized velocities, summing up and rearranging the expressions:
\begin{equation}
	\label{bilenmoltipl}
	\dfrac{d{\cal E}}{dt}
	=-\dfrac{\partial {\cal L}}{\partial t}+
	\sum\limits_{\nu=1}^k \sum\limits_{i=1}^n\lambda_\nu \dfrac{\partial \phi_\nu}{\partial {\dot q}_i} {\dot q}_i.
\end{equation}
	
\begin{equation}
	\label{bilenu}
	\dfrac{d {\cal E}^*}{dt}
	=
	-\dfrac{\partial {\cal L}^*}{\partial t}+
	\sum\limits_{\nu=1}^k ({\overline \alpha}_\nu-\alpha_\nu)\dfrac{\partial {\cal L}^*}{\partial q_{m+\nu}}
	+\sum\limits_{\nu=1}^k {\overline B}_\nu \dfrac{\partial T}{\partial {\dot q}_{m+\nu}}
\end{equation}
where ${\overline \alpha}_\nu$ is defined in (\ref{baralpha}) and (see also (\ref{b}))
\begin{equation}
	\label{barb}
	{\overline B}_\nu (q_1, \dots, q_n, {\dot q}_1, \dots, {\dot q}_m,t)
	=\sum\limits_{r=1}^m B_r^\nu {\dot q}_r
	=\dfrac{d}{dt}({\overline \alpha}_\nu-\alpha_\nu)-\sum\limits_{\mu=1}^k
	\dfrac{\partial \alpha_\nu}{\partial q_{m+\mu}}
	({\overline \alpha}_\mu -\alpha_\mu)+\dfrac{\partial \alpha_\nu}{\partial t}, \qquad \nu=1, \dots, k.
\end{equation}
Evidently, if the constraints (\ref{constr}) are absent (holonomic case), both (\ref{bilenmoltipl}) and (\ref{bilenu}) reduce to the standard energy balance for ${\cal L}^*={\cal L}$:

\begin{equation}
	\label{bilenollagr}
	\dfrac{d}{dt}\dfrac{\partial {\cal L}}{\partial {\dot q}_i}-\dfrac{\partial {\cal L}}{\partial q_i}=0\;\;\Rightarrow\;\;
	\dfrac{d}{dt}
	\left(\sum\limits_{i=1}^n {\dot q}_i \dfrac{\partial {\cal L}}{\partial {\dot q}_i}-{\cal L}\right)=
	-\dfrac{\partial {\cal L}}{\partial t}
\end{equation}

\section{The case ${\overline \alpha}_\nu=\alpha_\nu$}

\noindent
As it is exptected from the governing equations introduced above, the special situation when the functions in (\ref{constrexpl}) verify the condition
\begin{equation}
\label{baralphaalpha}
{\overline \alpha}_\nu=\alpha_\nu\quad \textrm{for any} \quad \nu=1,\dots, k 	
\end{equation}
where ${\overline \alpha}_\nu$ is defined in (\ref{baralpha}), deserves to be analyzed. 

\noindent
Many of the nonholonomic models studied in literature starting with the first examples fall into the category (\ref{baralphaalpha}): for istance, 
rigid bodies rolling without sliding on a plane or on a surface, 
knife edges or sleighs sliding on a horizontal plane, some joints simulating the basic functioning of 
a vehicle are formulated by linear homogeneous constraints of the type

	\begin{equation}
		\label{linom}
		\sum\limits_{i=1}^n\sigma_{\nu,i}(q_1,\dots, q_n,t){\dot q}_i=0 \qquad \nu=1, \dots, k
	\end{equation}
which verify (\ref{baralphaalpha}).

\noindent
Again, the kinematic constraints with homogeneous quadratic functions 
	\begin{equation}
		\label{quadrom}
		\sum\limits_{i,j=1}^n a_{i,j}^{(\nu)}(q_1, \dots, q_n,t){\dot q}_i {\dot q}_j=0, \qquad \nu=1, \dots, k
	\end{equation}
which include the conditions of parallel velocities (${\dot P}_1\wedge {\dot P}_2={\bf 0}$), orthogonal velocities (${\dot P}_1\cdot {\dot P}_2=0$) or same lenght ($|{\dot P}_1|=|{\dot P}_2|$), fulfill the request (\ref{baralphaalpha}).

\begin{rem}
Simple examples of constaints that do not verify (\ref{baralphaalpha}) are 
affine nonholonomic constraints of degree $p$, with $p$ positive integer:
\begin{equation}
	\label{constraffp}
	\sum\limits_{j=1}^n\sigma_{\nu,j}(q_1, \dots, q_n,t){\dot q}_j^p+\zeta_\nu(q_1, \dots, q_n, t)=0, \quad \nu=1, \dots, k
\end{equation}
The constraint on the magnitude of the velocity $|{\dot P}|=C(t)$, with $C(t)$ given nonnegative function, is part of (\ref{constraffp}) with $p=2$; for $p=1$ we have linear affine constraint. 
The explicit form (\ref{constrexpl}) is of the type
\begin{equation}
	\label{constrexplaff}
	\alpha_\nu= {\dot q}_{m+\nu}=(\pm 1)^{p+1}\left(\sum\limits_{j=1}^m \alpha_{\nu,j}(q_1, \dots, q_n,t){\dot q}_j^p+\beta_\nu(q_1, \dots, q_n,t)\right)^{1/p}, \qquad \nu=1, \dots, k
\end{equation}
for suitable coefficients $\alpha_{\nu,j}$ and $\beta_\nu$. The function (\ref{baralpha}) can be put in the form
$$
{\overline \alpha}_\nu= \alpha_\nu \dfrac{ \sum\limits_{i=1}^m\alpha_{\nu,i} {\dot q}_i^p}
{\sum\limits_{j=1}^m \alpha_{\nu,j}{\dot q}_j^p+\beta_\nu} \qquad \nu=1, \dots, k
$$
so that (\ref{baralphaalpha}) is valid if and only if  $\beta_\nu=0$, which means that the constraints are homogeneous.
\end{rem}

\noindent
Focusing now on the class of constraints (\ref{baralphaalpha}), at least three essential properties have to be pointed out, whenever (\ref{baralphaalpha}) holds:

\begin{itemize}
\item[$(1)$] the virtual displacements (\ref{deltar}), which may appear somehow artifical in their definition, reveal the physical meaning of settling along the virtual velocities, in the following sense: 
formally referring to the $N$--points system ${\bf r}_i(q_1, \dots, q_n,t)$ as in Remark 1, the velocity ${\dot {\bf r}_i}=\sum\limits_{j=1}^n\dfrac{\partial {\bf r}_i}{\partial q_r}{\dot q}_j + \dfrac{\partial {\bf r}_i}{\partial t}$ makes us consider the term
${\widehat {\dot {\bf r}}}_i=\sum\limits_{j=1}^n\dfrac{\partial {\bf r}_i}{\partial q_r}{\dot q}_j$ as the component consistent with the blocked configuration (virtual velocity). In the presence of (\ref{constrexpl}) one has	
\begin{equation}
	\label{velr}
	{\widehat {\dot {\bf r}}}_i=\sum\limits_{r=1}^m\dfrac{\partial {\bf r}_i}{\partial q_r}{\dot q}_r +
	\sum\limits_{\nu=1}^k \dfrac{\partial {\bf r}_i}{\partial q_{m+\nu}}\alpha_\nu=
\left(	\sum\limits_{r=1}^m\dfrac{\partial {\bf r}_i}{\partial q_r}
+\sum\limits_{\nu=1}^k 
\dfrac{\partial {\bf r}_i}{\partial q_{m+\nu}}\dfrac{\partial \alpha_\nu}{\partial {\dot q}_r}\right) 
{\dot q}_r, \qquad i=1, \dots, N
\end{equation}
where the last equality is due to (\ref{baralphaalpha}). On the other hand, the calculation of (\ref{deltar}) leads to 
$$
\delta {\bf r}_i = \sum\limits_{j=1}^n \dfrac{\partial {\dot {\bf r}}_i}{\partial  {\dot q}_j}\delta q_j=
\sum\limits_{r=1}^m \left( \dfrac{\partial {\bf r}_j}{\partial q_r}
+\sum\limits_{\nu=1}^k \dfrac{\partial {\bf r}_i}{\partial q_{m+\nu}}\dfrac{\partial \alpha_\nu}{\partial {\dot q}_r}\right) \delta q_r,
 \quad i=1, \dots, N
$$
where (\ref{virtdispdip}) has been taken into account. Hence, we see that, a part from the formal appearance through $\delta q_r$ or ${\dot q}_r$, the two vectors display the same direction.

\item[$(2)$] At the same time, the vanishing $\sum\limits_{i=1}^n {\cal R}^{(i)}\delta q_i =0$ (see (\ref{riphinu})) does actually represent the absence of virtual work $\sum\limits_{i=1}^n {\cal R}^{(i)}{\dot q_i} =0$: a wat to check that goes through the  relation 
$$		
\sum\limits_{i=1}^n{\cal R}^{(i)} {\dot q}_i = 
\sum\limits_{\nu=1}^k
(\alpha_\nu-{\overline \alpha}_\nu)
\left(\dfrac{d}{dt}\dfrac{\partial {\cal L}}{\partial {\dot q}_{m+\nu}}-\dfrac{\partial {\cal L}}{\partial q_{m+\nu}}\right)
$$
which can be deduced from (\ref{eqlagrl}).

\item[$(3)$] The relation (\ref{rele}) shows that (\ref{e}) and (\ref{estar}) overlap, that is calculating the energy by considering the Lagrangian function with all the velocities ${\dot q}_i$, $i=1, \dots, n$ or the restricted function (\ref{lstar}) of the independent velocities only ${\dot q}_r$, $r=1, \dots, m$, leads to the same result.
\item[$(4)$] The balance equation (\ref{bilenu}) reduces to (see also (\ref{barb}))
\begin{equation}
	\label{bilena}
\dfrac{d {\cal E}^*}{dt}
=-\dfrac{\partial {\cal L}^*}{\partial t}
+\sum\limits_{\nu=1}^k \dfrac{\partial \alpha_\nu}{\partial t} \dfrac{\partial T}{\partial {\dot q}_{m+\nu}}
=-\dfrac{\partial {\cal L}}{\partial t}
\end{equation}
and it reveals the first integral of motion ${\cal E}={\cal E}^*$, whenever the Lagrangian function ${\cal L}$ does not depend explicitly on time $t$.
\end{itemize}

\begin{rem}
The second equality in (\ref{bilena}) shows that the energy may be conserved even if the constraints (\ref{constr}) depend explicitly on time $t$ (rheonomic constraints): a simple example is the motion of a point $P$ of mass $M$ whose velocity has to at any time the direction of $\overrightarrow{PQ}$, where the motion of $Q$ is assigned by the functions $x_Q(t), y_Q(t),z_Q(t)$ (pursuing motion). 
The constraints (\ref{constrexpl}) are 
$$
		\left\{
		\begin{array}{ll}
			{\dot q}_2 =\dfrac{y_Q(t)-q_2}{x_Q(t)-q_1}{\dot q}_1=\alpha_{1,1}(q_1, q_2, t){\dot q}_1, &
			\\
			{\dot q}_3 =\dfrac{z_Q(t)-q_3}{x_Q(t)-q_1}{\dot q}_1=\alpha_{2,1}(q_1, q_3, t){\dot q}_1
		\end{array}
		\right.
$$
and they verify (\ref{baralphaalpha}), as it can be easily checked. If there are not active forces, one has
${\cal L}=T=\frac{1}{2}M({\dot q}_1^2+{\dot q_2}^2+{\dot q}_3^2)$ and 
$$
{\cal E}={\cal E}^*
=\dfrac{1}{2}M{\dot q}_1^2 \left(1+
		\dfrac{(q_2-\eta(t))^2+(q_3-\zeta(t))^2}{(q_1-\xi(t))^2} \right).
$$
is conserved by virtue of (\ref{bilena}). This means that $|{\dot P|}$ is constant during the motion.
\end{rem}

\noindent
The unifying and physically expressive role of (\ref{baralphaalpha}) motivates the following mathematical investigation.

\noindent
As it occurs in the holonomic case, it is reasonable to expect that the same set of nonholonomic restriction may be formulated by more than one set of constraint equations: a simple instance is the following

\begin{exe}
Consider a system of two points $P_1$ and $P_2$ constrained in a way that their velocities are perpendicular to the straight line joining them: ${\dot P}_1\cdot \overrightarrow{P_1P_2}=0$, ${\dot P}_2\cdot \overrightarrow{P_1P_2}=0$; referring to (\ref{constr}), the condition is formulated as
$$
\left\{
\begin{array}{l}
(x_1-x_2){\dot x}_1+(y_1-y_2){\dot y}_1=0\\
\\
(x_1-x_2){\dot x}_2+(y_1-y_2){\dot y}_2=0  
\end{array}
\right.
$$
(we keep the cartesian coordinates for clarity). 
Calling $B$ the midpoint of the segment $P_1P_2$, 
the same effect can be achieved by imposing that the distance between the points is constant and 
the velocity of the midpoint is orthogonal to the joining line, namely the conditions 
$|\overrightarrow{P_1P_2}|=\ell>0$, ${\dot B} \cdot \overrightarrow{P_1P_2}={\bf 0}$.
Actually, the corresponding system in terms of cartesian coordinates
$$
\left\{
\begin{array}{l}
	(x_1-x_2)({\dot x}_1-{\dot x}_2)+(y_1-y_2)({\dot y}_1-{\dot y}_2)=0\\
\\
	(x_1-x_2)({\dot x}_1+{\dot x}_2)+(y_1-y_2)({\dot y}_1+{\dot y}_2)=0 
\end{array}
\right.
$$
is evidently equivalent to the one written just above (the first constraint expresses the invariable distance in the differential form).

\noindent
Further equivalent conditions can be provided, but attention must be paid to the fact that the equivalence could be lost in correspondence of some particual motions: concerning this example, the pair of conditions 
${\dot B} \cdot \overrightarrow{P_1P_2}={\bf 0}$, ${\dot P}_1\wedge {\dot P}_2={\bf 0}$ (parallel velocities) 
is equivalent to the previous ones whenever the velocity ${\dot B}$ is not null (the critical motions are the rotations around $B$).
Again, the combination of $|\overrightarrow{P_1P_2}|=\ell>0$, ${\dot P}_1\wedge {\dot P}_2={\bf 0}$ produces the same effect if ${\dot P}_1\not = {\dot P}_2 $
(the critical motions are of translations).
The analysis of the Jacobian matrices of the various cartesian systems confirms the statements of the Remark without difficulty.
\end{exe}

\begin{rem}
In order to find equivalent sets of kinematic conditions the constraint equations are not necessarily linear with respect to the velocities as in the previous case: an example can be formulated by combining two of the three conditions
$|\overrightarrow{P_1P_2}|=\ell>0$,  ${\dot B}\wedge \overrightarrow{P_1P_2}={\bf 0}$, $|{\dot P}_1|=|{\dot P}_2|$ (same magnitude of the velocities) corresponding respectively to the linear and nonlinear equations
\begin{equation}
	\label{rem6}
\begin{array}{l}
(x_1-x_2)({\dot x}_1-{\dot x}_2)+(y_1-y_2)({\dot y}_1-{\dot y}_2)=0, \quad 
({\dot x}_1+{\dot x}_2)({\dot x}_1-{\dot x}_2)+({\dot y}_1+{\dot y}_2)({\dot y}_1-{\dot y}_2)=0, \\ 
(x_1-x_2)({\dot y}_1+{\dot y}_2)-(y_1-y_2)({\dot x}_1+{\dot x}_2)=0.
\end{array}
\end{equation}
\end{rem}

\subsection{The mathematical aspect}

\noindent
From the mathematical point of view it is quite simple to understand which functions verify the condition
(\ref{baralphaalpha}). 
The context that emerges is that of homogeneous functions: we start by reminding that a function $f(\xi_1, \dots, \xi_n)$ defined on a domain ${\cal D}\subseteq {\Bbb R}^n$ is a positive homogeneous function of degree $\sigma \in {\Bbb R}$ if
\begin{equation}
	\label{hom}
	f(\lambda \xi_1, \dots, \lambda \xi_\ell)=\lambda^{\sigma}	f(\xi_1, \dots,\xi_\ell)\quad \forall\;(\xi_1, \dots, \xi_\ell)\in {\cal D}\;\;\textrm{and}\;\;\forall\;\lambda>0.
\end{equation}
The functions (\ref{hom}), when differentiable, are distinguished by the following 
\begin{propr} (Euler's homogeneous function theorem)  
	A function $f\in {\cal C}^1({\cal D})$ is a positive homogeneous function of degree $\sigma$ if and only if 
\begin{equation}
		\label{eulero}
		\sum\limits_{i=1}^\ell \xi_i\dfrac{\partial f}{\partial \xi_i}(\xi_1, \dots, \xi_\ell)= \sigma 
		f(\xi_1, \dots, \xi_\ell)\quad \forall\;(\xi_1, \dots, \xi_\ell)\in {\cal D}.
\end{equation}
\end{propr}

\noindent
If we differentiate (\ref{hom}) with respect to $\xi_i$, we run into the following consequence:

\begin{propr} 
If a function $f(\xi_1, \dots, \xi_n)\in {\cal C}^1({\cal D})$, ${\cal D}\subseteq {\Bbb R}^\ell$ is a positive homogeneous function of degree $\sigma>0$, then each derivative $\dfrac{\partial f}{\partial x_i}$, $i=1, \dots, n$, is a positive homogeneous function of degree $\sigma-1$.
\end{propr}

\noindent
A further property that we use immediately after and that is easy to verify is the 
\begin{propr}
If $F_1$, $F_2$ are two homogeneous functions of degree $\sigma_1$ and $\sigma_2$ respectively, then the product $F_1F_2$ is a homogeneous function of degree $\sigma_1+\sigma_2$, the ratio $F_1/F_2$ (where defined) is a homogeneous function of degree $\sigma_1-\sigma_2$.
\end{propr}

\noindent
The main result of our mathematical digression is the following
\begin{prop}
Consider a system of $k$ equations
$$
\left\{
\begin{array}{l}
	F_1(\xi_1, \dots, \xi_\ell, \eta_1, \dots, \eta_k)=0 \\
\qquad 	\dots \quad \dots \quad \dots \\
\qquad 	\dots \quad \dots \quad \dots \\	
	F_k(\xi_1, \dots, \xi_\ell, \eta_1, \dots, \eta_k)=0
\end{array}
\right.
$$
where each $F_\nu\in {\cal C}^1 ({\cal D})$, ${\cal D}\subseteq {\Bbb R}^{\ell+k}$ is a homogeneous function of degree $\sigma_\nu$, $\nu=1, \dots, k$, that is 
\begin{equation}
	\label{constrgen}
F_\nu(\lambda \xi_1, \dots, \lambda \xi_\ell, \lambda y_1, \dots, \lambda y_k)=\lambda^{\sigma_\nu}	
F(\xi_1, \dots,\xi_\ell, y_1, \dots, y_k)\quad\;\lambda>0
\end{equation}
in any point of $D$. Then, assuming that the set of zeros can be written by means of the $k$ implicitly 
defined functions 
$$
\left\{
\begin{array}{l}
y_1=\psi_1(\xi_1, \dots, \xi_\ell) \\
	\qquad 	\dots \quad \dots \quad \dots \\
	\qquad 	\dots \quad \dots \quad \dots \\	
y_k=\psi_k(\xi_1, \dots, \xi_\ell)
\end{array}
\right.
$$	
the functions $\psi_1$, $\dots$, $\psi_k$ turn out to be positive homogeneous functions of degree $1$.
\end{prop}

\noindent
{\bf Proof}. Since $F_\nu$ is a homogeneous function of degree $\sigma_\nu$, (\ref{eulero}) implies 
$$
\dfrac{\partial F_\nu}{\partial \xi_1}\xi_1+\,\dots\,+\dfrac{\partial F_\nu}{\partial \xi_\ell}\xi_\ell+
\dfrac{\partial F_\nu}{\partial \eta_1}\eta_1+\,\dots\,+\dfrac{\partial F_\nu}{\partial \eta_k}\eta_k=
\sigma_\nu F_\nu(\xi_1, \dots,\xi_\ell, y_1, \dots, y_k), \quad \nu=1, \dots, k
$$
The calculation of $\eta_1$, $\dots$, $\eta_k$ from the previous identities consists in solving a $k\times k$ linear system whose solutions are
$$
\eta_1=\dfrac{1}{D}
\left| 
\begin{array}{llll}     
	\sigma_1 F_1 -\sum\limits_{r=1}^m \dfrac{\partial F_1}{\partial \xi_r}\xi_r& 
	 \dfrac{\partial F_1}{\partial \eta_2} & \dots & \dfrac{\partial F_1}{\partial \eta_k} \\
\quad \dots \dots \dots & \dots & \dots & \dots \\	 
	\sigma_k F_k -\sum\limits_{r=1}^m \dfrac{\partial F_k}{\partial \xi_r}\xi_r& 
\dfrac{\partial F_k}{\partial \eta_2} & \dots & \dfrac{\partial F_k}{\partial \eta_k} 
\end{array}  
\right|
\qquad \dots \qquad 
\eta_k=\dfrac{1}{D}
	\left| 
	\begin{array}{llll}     
		\dfrac{\partial F_1}{\partial \eta_1} & \dots & \dfrac{\partial F_1}{\partial \eta_{k-1}} 
	&	\sigma_1 F_1 -\sum\limits_{r=1}^m \dfrac{\partial F_1}{\partial \xi_r}\xi_r
		\\
		\dots & \dots & \dots & \quad \dots \dots \dots \\	 
		\dfrac{\partial F_k}{\partial \eta_1} & \dots & \dfrac{\partial F_k}{\partial \eta_{k-1}} 
	& 	\sigma_k F_k -\sum\limits_{r=1}^m \dfrac{\partial F_k}{\partial \xi_r}\xi_r
	\end{array}  
	\right|
$$
where the vertical bars stand for the determinant of the contained matrix and 
$D=	
\left| 
\begin{array}{lll}     
	\dfrac{\partial F_1}{\partial \eta_1} & \dots & \dfrac{\partial F_1}{\partial \eta_k} \\
	\dots & \dots & \dots \\	 
	\dfrac{\partial F_k}{\partial \eta_1} & \dots & \dfrac{\partial F_k}{\partial \eta_k} 
\end{array}  
\right|
$.

\noindent
Owing to Property 2, each derivative $\dfrac{\partial F_\nu}{\partial \eta_\mu}$, $\nu, \mu=1, \dots, k$ is a homogeneous function of degree $\sigma_\nu-1$; the Leibniz formula for the determinant prescribes the algebric sum of terms of the type $\prod\limits_{i,j=1}^k \dfrac{\partial F_{\nu_i}}{\partial {\mu_j}}$, where the suffixes $\nu_i$ are all different. Hence, by virtue of Property 3 and considering that the sum of homogeneous functions with common degree is evidently a homogeneous function of same degree, we have that 
the determinant $D$ is a homogeneous function of degree 
$(\sigma_1-1)+(\sigma_2-1)+\dots+(\sigma_k-1)=\sum\limits_{\nu=1}^k \sigma_\nu-k$.
The same argument applies to the matrices at the numerator of $\eta_\nu$, whose determinants turn out to be  homogeneous function of degree 
$\sum\limits_{\nu=1}^k \sigma_\nu-(k-1)$ (indeed the entries of the $\nu$--th column of the matrix pertinent to $\eta_\nu$ are homogeneous function of degree $\sigma_\nu$).
Invoking again Property 3 this time for the ratio, we overall obtain that each $\eta_\nu$ is a homogeneous function of degree $\sum\limits_{\nu=1}^k \sigma_\nu-(k-1)- \left(\sum\limits_{\nu=1}^k \sigma_\nu-k\right) =1$.
$\quad\square$

\subsection{The physical application}

\noindent
If we compare the property (\ref{eulero}) with the request (\ref{baralphaalpha}), 
we recognise in the latter one the characteristic condition for the function $\alpha_\nu$ to be a homogeneous function of degree $1$ with respect to the variables ${\dot q}_r$: thus, setting $\ell=m$ and ${\dot q}_r=\xi_r$, $r=1, \dots, m$, (\ref{baralpha}), (\ref{hom}) and (\ref{eulero}) entail the following

\begin{prop}
The function $\alpha_\nu$, $\nu=1, \dots, k$, verifies ${\overline \alpha}_\nu=\alpha_\nu$ if and only if $\alpha_\nu$ is a positive homogeneous function of degree $1$ with respect to the kinetic variables ${\dot q}_1$, $\dots$, ${\dot q}_m$, namely if and only if 
	\begin{equation}
		\label{alphahom}
		\alpha_\nu (q_1, \dots, q_n, \lambda {\dot q}_1, \dots, \lambda {\dot q}_m, t)=
		\lambda \alpha_\nu (q_1, \dots, q_n, {\dot q}_1, \dots, {\dot q}_m, t)\quad \textrm{for any}\;\;\lambda>0.
	\end{equation}
\end{prop}
A more general and noteworthy result is provided by Proposition 1, which refers directly to the structure of the assigned constraint functions and not to the implicitly defined ones: actually, with respect to (\ref{constrgen}) we consider the kinetic variables ${\dot q}_i$ of (\ref{constr}) as ${\dot q}_r=\xi_r$, $r=1, \dots, m=\ell$ and ${\dot q}_{m+\nu}=\eta_\nu$, $\nu=1, \dots, k$ and we make use of Proposition 
$1$ for the following application:

\begin{prop}
If $\phi_\nu(q_1, \dots, q_n, {\dot q}_1, \dots, {\dot q}_n,t)$ of (\ref{constr}) are homogeneous functions (even with different degrees) of the generalized velocities ${\dot q}_1$, $\dots$, ${\dot q}_n$, then however the explicit forms (\ref{constrexpl}) are chosen, they satisfy condition (\ref{baralphaalpha}).
\end{prop}

\noindent
As we have already highlighted, the same system can be treated either with linear kinematic constraints or with nonlinear kinematic constraints (or a mixture of them):
this fact does not affect the definition and conservation of the energy of the system, if the latter falls into the category admitted by the previous Proposition.

\noindent
For instance, the system we considered in (\ref{rem6}) is described by homogeneous functions, whichever pair of constraints is choosen, except for any points where equivalence is lost, as we remarked in Example 1. In any case, the explicit functions (\ref{constrexpl}) will verify condition (\ref{baralphaalpha}).

\noindent
Among many other examples we can present, we add the following nonholonomic system, studied in \cite{zekint}.

\begin{exe}
Consider on a vertical plane two material points $P_1$, $P_2$ of mass $M_1$, $M_2$ respectively: let us call attention on the three kinematic restrictions

	\begin{itemize}
		\item[$(i)$] the velocities are perpendicular: ${\dot P}_1\cdot {\dot P}_2=0$,
		\item[$(ii)$]  the velocity of one of them is perpendicular to the straight line joining the points: ${\dot P}_1\cdot \overrightarrow{P_1P_2}=0$, 
		\item[$(iii)$]  the velocity of the other point is parallel to the joining line: 
		${\dot P}_2\wedge \overrightarrow{P_1P_2}={\bf 0}$.
	\end{itemize}
	
\noindent
By formulating the conditions as 
	$$
\begin{array}{lll}
{\dot x}_1{\dot x}_2+{\dot y}_1{\dot y}_2=0, & (x_1-x_2){\dot x}_1+(y_1-y_2){\dot y}_1=0, & 
(x_1-x_2){\dot y}_2-(y_1-y_2){\dot x}_2=0
	\end{array}
	$$
respectively, it can be easily seen that they are not independent and two of them imply the third with the exceptions noted alongside (the explication for the excluded configurations is clear):
	$$
	\begin{array}{lll}
		(i), \, (ii)\;\Rightarrow\; (iii)\;\;\textrm{if}\;\;{\dot P}_1\not = {\bf 0}, & 
		(i), \, (iii) \;\Rightarrow \;(ii)\;\;\textrm{if}\; \; {\dot P}_2\not = {\bf 0}, &
		(ii), \, (iii) \;\Rightarrow \;(i)\;\;\textrm{if}\; \;P_1\not \equiv P_2.
	\end{array}
	$$
\end{exe}

\noindent
Even admitting that the velocities can vanish but excluding the overlapping of the points, we opt for the latter possibility and write (\ref{constrexpl}) as
$$
\left\{
\begin{array}{l}
	{\dot q}_3 =\alpha_1 (q_1, q_2, q_3, q_4, {\dot q}_1)=-\alpha {\dot q}_1 \\
	\\
	{\dot q}_4 =\alpha_2 (q_1, q_2, q_3, q_4, {\dot q}_2)=\dfrac{1}{\alpha}{\dot q}_2
\end{array}\right.
$$
where we have set $(q_1, q_2, q_3, q_4)=(x_1, x_2, y_1, y_2)$ and defined $\alpha (q_1, q_2, q_3, q_4)=\dfrac{q_1-q_2}{q_3-q_4}$. 
The assumption (\ref{baralphaalpha}) holds because the constraints are part of (\ref{linom}).
Assuming that an internal elastic force of constant $\kappa$ and the weight are acting, the function (\ref{lstar}) is 
$$
{\cal L}^*(q_1, q_2, q_3, q_4, {\dot q}_1, {\dot q}_2)=\dfrac{1}{2}M_1 \left(1+\alpha^2\right){\dot q}_1^2+\dfrac{1}{2}M_2\left(1+\dfrac{1}{\alpha^2}\right){\dot q}_2^2-\dfrac{1}{2}\kappa(q_3-q_4)^2(1+\alpha^2) -M_1q_3-M_2q_4
$$
where the terms concerning $U$ are clear. Equation (\ref{bilena}) entails the conservation of the quantity (\ref{estar}), coinciding in this case with the energy of the system ${\cal E}=T-U$.

\section{Conclusion and next investigation}

\noindent
For nonholonomic systems verifying (\ref{baralphaalpha}), the formulation of the motion through the virtual displacements provided by the ${\check {\rm C}}$hetaev condition (\ref{cetaev}) appears
natural and not devoid of physical meaning. This represents a convincing extension of the holonomic case equipped by a well-established and dated theory.
The absence of a justification of the condition (\ref{cetaev}) starting from the laws of mechanics
sometimes claimed in literature, is overcome when the kinematic constraints are of the type (\ref{baralphaalpha}).

\noindent
As we have already highlighted, for these systems the notion of virtual displacement assumes a physical sense and also the energy of the system has a correct meaning, as for holonomic systems.
Likewise to the latter ones, when the Lagrangian function does not depend on time the energy is conserved, as 
(\ref{bilena}) states.

\noindent
To understanding which functions verify (\ref{baralphaalpha}) is simple, since it is a plain implementation of Euler's theorem on homogeneous functions.
The further step we performed was to indicate a large class of constraints for which the explicit form (\ref{constrexpl}) shows the functions $\alpha_\nu$ of the category (\ref{baralphaalpha}): this occurs for the constraints formulated by homogeneous functions (with not necessarily the same degree) of the generalized velocities ${\dot q}_i$, $i=1, \dots, n$. 

\noindent
The incoming investigation will concern a sort of inverted question: if one has to do with explicit functions $\alpha_\nu$ verifiying (\ref{baralphaalpha}), necessarily the originating functions $\phi_\nu$ which define them are homogeneous functions in the kinematic variables? This is not an irrelevant point, since the category of nonholonomic constraints satisfying the physical properties $(1)$--$(4)$ listed just after Remark 4 would become entirely and clearly defined, as the systems subject to homogeneous constraints, i.~e.~the ${\check {\rm C}}$etaev condition would be completely legitimated for this category of systems.
In other words, the constrained systems for which the condition acquires a physical meaning and extends in a natural way the virtual displacements of the holonomic systems are presumed to be all and only those which admit at least one set of conditions (\ref{constr}) formulated with equations homogeneous in the velocities.

\noindent
However, the mathematical aspect is anything but straightforward: as it incidentally emerged in some passage of the script, the geometric and kinematic restrictions can be implemented in more than a way, so that different lists of functions (\ref{constr}) can formulate the 
same system. Then, from the mathematical point of view, the issue consists in investigating the zero set defined by (\ref{constr}) and wonder if for each explicit set $\alpha_\nu$, $\nu=1, \dots, k$ a list of generating functions $\phi_\nu$ which are homogeneous in the velocities can always be 
found.


\begin{thebibliography}{10}
	
\bibitem{appell} Appell P.~, Sur les liasons exprim\'ees par des relations non lin\'eaires entre les vitesses, C.~R.~Acad.~Sci.~Paris {\bf 152}, 1197--319, 1200, 1911.
	
\bibitem{bloch} Bloch, A.~M.~, Mardsen, J.~E.~and Zenkov, D.~V.~, Quasivelocities and Symmetries in Non-Holonomic Systems, Dynamical Systems {\bf 24}, 187--222, 2009.
	
\bibitem{cron} Cronstr{\" o}m, C.~, On the Compatibility of Nonholonomic Systems and Related Variational Systems, Acta Physica Universitatis comenianae Vol.~L--LI, 1 and 2, 25--36, 2010.
	
\bibitem{cetaev} ${\check {\rm C}}$etaev, N.~G.~, On the Gauss Principles, Papers on Analytical Mechanics {\bf 323}, Science Academy, 1962.
	
\bibitem{fasso} Fass\`o, F.~, Sansonetto, N.~, Conservation of Energy and Momenta in Nonholonomic Systems with Affine Constraints, Regular and Chaotic Dynamics {\bf 20} n.~4, 449--462, 2015.
	
\bibitem{flan} Flannery M.~R.~, The enigma of nonholonomic constraints, American Journal of Physics {\bf 73}, 265--272, 2005.
	
\bibitem{leon1}  de Le\'on M.~, A historical review on nonholonomic mechanics, Revista de la Real Academia de Ciencias Exactas, F\'isicas y Naturales. Serie A, Matem\'aticas {\bf 106}(1), 2012.
	
\bibitem{leon2} de Le\'on, M.~, de Diego, D.~M.~, On the geometry of non-holonomic Lagrangian systems, Mechanical systems with nonlinear constraints, J.~Math.~Phys.~{\bf 37}, 3389--3414, 1996.
	
\bibitem{li} S. M. Li and J. Berakadar: A generalization of the Chetaev condition for nonlinear nonholonomic constraints: The velocity-determined virtual displacement approach, Reports on Mathematical Physics	
{\bf 63} Issue 2, 179--189, 2009.
	
\bibitem{marle} Marle, C--M, Various approaches to conservative and nonconservative nonholonomic systems
Reports on Mathematical Physics {\bf 42}, Issues 1–-2, 211--229, 1998.
	
\bibitem{massa} Massa, E.~, Pagani, E.~, Classical Dynamics of non--holonomic systems: a geometric approach, Ann.~Inst.~H.~Poincar\'e Phys.~Th\'eor.~{\bf 55}, 511--544, 1991.
	
\bibitem{neimark} Ne${\check {\rm i}}$mark Ju.~I.~, Fufaev N.~A.~, Dynamics of Nonholonomic Systems, Providence: American Mathematical Society, Translations of Mathematical Monographs {\bf 33}, 1972.
	
\bibitem{papa} J. G. Papastravidis, J.~G.~, Time--integral Variational Principles for Nonlinear Nonholonomic Systems, J.~Appl.~Math.~{\bf 64}, 985--991, 1997.
	
	
\bibitem{ranada} Cari\~{n}ena, J.~F.~, Ra\~{n}ada, M.~F.~, Lagrangian systems with constraints: a geometric approach to the method of Lagrange multipliers, J.~Phys.~A: Math.~Gen.~{\bf 26}, 1335-1351, 1993.
	
\bibitem{rumy} V. V. Rumyantsev, V.~V.~, Forms of Hamilton's principle for nonholonomic systems, Mechanics, Automatic and Robotics {\bf 2}, n.~10, 1035--1048, 2000.
		
		
\bibitem{rumycet} V. V. Rumyantsev, V.~V.~, On the Chetaev principle (In Russian), Dokl.~Akad.~Nauk SSSR {\bf 210}  n.~4, 787--790, 1973.
	
\bibitem{saleh} Salehani, M.~K.~, A Jet Bundle Approach to the Variational Structure of Nonholonomic Mechanical Systems, Reports on Mathematical Physics {\bf 83} Issue 3, 373--385, 2019.
		
\bibitem{tal} Talamucci, F., Rheonomic Systems with Nonlinear Nonholonomic Constraints: The Voronec Equations, Regular and Chaotic Dynamics {\bf 25} n.~6, 662--673, 2020.
	
\bibitem{zekovic1} Zekovi\'c, D.~N.~,  Dynamics of mechanical systems with nonlinear nonholonomic constraints – I The history of solving the problem of a material realization of a nonlinear nonholonomic constraint, Z.~Angew.~Math.~Mech.~{\bf 91} n.~11, 883--898, 2011.
	
\bibitem{zekint}  Zekovi\'c, D.~N.~, Linear integrals of non--holonomic systems with non--linear constraints, Journal of Appl.~Math.~and Mech.~{\bf 69}, 832--836, 2005.
\end{thebibliography}
\end{document}